\documentclass[lettersize,journal]{IEEEtran}
\usepackage{amsmath,amssymb,amsfonts}%
\usepackage{array}

\usepackage{textcomp}
\usepackage{stfloats}
\usepackage{url}
\usepackage{verbatim}
\usepackage{graphicx}
\usepackage{cite}
\usepackage{algorithmic}
\usepackage{subfigure}
\usepackage[utf8]{inputenc}
\usepackage[T1]{fontenc}

\usepackage{multirow}%
\usepackage{amsthm}%
\usepackage{mathrsfs}%
\usepackage{xcolor}%
\usepackage{textcomp}%
\usepackage{manyfoot}%
\usepackage{booktabs}%
\usepackage{algorithm}%
\usepackage{listings}%
\usepackage{hyperref}
\usepackage{caption}
\makeatletter

\newcommand{\Rmnum}[1]{\expandafter\@slowromancap\romannumeral #1@}
\makeatother
\begin{document}

\title{Multi-Branch DNN and CRLB-Ratio-Weight Fusion for
Enhanced DOA Sensing via a Massive H$^2$AD MIMO Receiver}

\author{Feng Shu,~\emph{Member},~\emph{IEEE}, Jiatong Bai, Di Wu, Wei Zhu, Bin Deng, Fuhui Zhou,~\emph{Senior Member},~\emph{IEEE}, and Jiangzhou Wang,~\emph{Fellow},~\emph{IEEE}

\thanks{Corresponding authors: Jiatong Bai, Di Wu, Wei Zhu and Feng Shu}
\thanks{Feng Shu is with the School of Information and Communication Engineering and Collaborative Innovation Center of Information Technology, Hainan University, Haikou 570228, China, and also with the School of Electronic and Optical Engineering, Nanjing University of Science and Technology, Nanjing 210094, China. (e-mail: shufeng0101@163.com).}
\thanks{Jiatong Bai and Bin Deng are with the School of Information and Communication Engineering, Hainan University, Haikou 570228, China. (e-mail: 18419229733@163.com; d2696638525@126.com).}
\thanks{Di Wu is with the ICAP Lab, School of Electronic Science and Technology, Hainan University, Haikou 570228, Hainan, China, and also with the PRISMA Lab, Department of Electrical Engineering and Information Technology, University of Naples Federico II, 80125 Naples, Italy.  (e-mail: hainuwudi@hainanu.edu.cn).}
\thanks{Wei Zhu is with the School of Energy and Power Engineering, Nanjing University of Science and Technology, Nanjing 210094, China.  (e-mail: zhuwei@njust.edu.cn).}
\thanks{Fuhui Zhou is with the College of Electronic and Information Engineering, Nanjing University of Aeronautics and Astronautics, Nanjing 210000, China. (email: zhoufuhui@ieee.org).}
\thanks{Jiangzhou Wang is with the National Mobile Communications Research Laboratory, Southeast University, Nanjing 210096, China, and also with the Pervasive Communication Research Center, Purple Mountain Laboratories,Nanjing 211111, China (e-mail: j.z.wang@seu.edu.cn).}
}



\maketitle

\begin{abstract}
As a green MIMO structure, massive H$^2$AD is viewed as a potential technology for the future 6G wireless network. For such a structure, it is a challenging task to design a low-complexity and high-performance fusion of target direction values sensed by different sub-array groups with fewer use of prior knowledge. To address this issue,
a lightweight Cramer-Rao lower bound (CRLB)-ratio-weight fusion (WF) method is proposed, which approximates inverse CRLB of each subarray using antenna number reciprocals to eliminate real-time CRLB computation. This reduces complexity and prior knowledge dependence while preserving fusion performance.
Moreover, a multi-branch deep neural network (MBDNN) is constructed to further enhance direction-of-arrival (DOA) sensing by leveraging candidate angles from multiple subarrays. The subarray-specific branch networks are integrated with a shared regression module to effectively eliminate pseudo-solutions and fuse true angles.  
Simulation results show that the proposed CRLB-ratio-WF method achieves DOA sensing performance comparable to CRLB-based methods, while significantly reducing the reliance on prior knowledge. More notably,
the proposed MBDNN has superior performance in low-SNR ranges. At SNR $= -15$ dB, it achieves an order-of-magnitude improvement in estimation accuracy compared to CRLB-ratio-WF method.
\end{abstract}

\begin{IEEEkeywords}
DOA, massive H$^2$AD MIMO, weighted fusion, multi-branch deep neural network.
\end{IEEEkeywords}

\section{Introduction}
\IEEEPARstart{M}{assive} or ultra-massive full-digital (FD) multiple-input multiple-output (MIMO) technology is anticipated to play an indispensable role in forthcoming sixth-generation (6G) communication systems \cite{wang2022vision, zhang20196g, 10810300, 9132710}, owing to its ability to deliver ultra-low latency and wide-area coverage.
In particular, FD-MIMO architectures based on uniform linear arrays (ULAs) offer high-precision direction-of-arrival (DOA) sensing. However, as the number of antennas grow to massive or ultra-massive levels, FD-MIMO systems are increasingly constrained by the so-called “three highs”: high hardware cost, high system complexity, and high energy consumption. These challenges significantly hinder its practical deployment and widespread application.

To tackle the aforementioned challenges, hybrid analog-digital (HAD) MIMO architectures have emerged as a promising solution, widely adopted to alleviate hardware complexity and reduce energy consumption \cite{zhang2021direction, abdelbadie2024doa, 9748022, 10460419}. 
In HAD systems, accurate DOA estimation plays a crucial role in beamforming\cite{9971745}, hybrid precoding design\cite{10689354}, AOA location\cite{7097001}, and channel estimation\cite{cui2022channel}, etc. Currently, the dominant HAD architectures include: fully connected HAD (FC-HAD), sub-connected HAD (SC-HAD), and switches-based HAD (SE-HAD) structures. Unlike traditional FD architectures where each antenna is connected to a dedicated radio frequency (RF) chain, the HAD architecture connects multiple antennas to a single RF chain via phase shifters (PS) or switches, thereby significantly reducing hardware cost.
In the FC-HAD architecture, each antenna is connected to all RF chains via individual analog PS. While this configuration provides high estimation accuracy, it suffers from high circuit complexity. The SC-HAD architecture partitions the antenna array into several subarrays, each connected to a dedicated RF chain. Therefore, under the same number of RF chains, the SC-HAD architecture has lower hardware cost than the FC-HAD structure. However, this cost and complexity reduction comes at the expense of a higher susceptibility to phase ambiguity. In contrast, the SE-HAD architecture employs switches to dynamically configure the connection paths between antennas and RF chains, thereby providing enhanced flexibility in energy control.

Although the SE-HAD architecture eliminates the need for phase shifters—resulting in the lowest hardware cost among the three structures—its performance is the worst, as each switch can connect to only one antenna at a time.
For the above three architectures, the existing research has explored the integration of deep learning (DL) \cite{wang2025, 10409285, 10323377} and optimization algorithms\cite{shen2024effective, 9594744, 9405998} to enhance DOA sensing performance. For example, a multi-layer deep neural network (DNN) was developed for the FC-HAD-UCA architecture to perform two-stage DOA estimation \cite{hu2019low}, where the first-stage DNN produced several coarse estimates that were subsequently refined by a second-stage DNN to obtain the final result.
In \cite{zhang2021direction}, a dynamic phase-shifting method based on the maximum likelihood criterion was proposed, which is compatible with all three mainstream HAD architectures, though it suffers from high computational complexity.
Li et al. \cite{li2023deep} developed a convolutional neural network (CNN) integrated with a noise-suppression module to process signal covariance matrices in overlapped HAD arrays, showcasing the robustness of DL in complex channel environments.
The SC-HAD architecture, which strikes a balance between cost and complexity, has gained attention for its ability to achieve high DOA estimation accuracy with relatively low hardware cost and system complexity, as demonstrated in \cite{shuHADDOA2018tcom}.

Despite recent advances, existing HAD architectures still encounter significant bottlenecks in practical deployment \cite{heath2016overview, chen2021fast,10410579}. Specifically, FC-HAD and SE-HAD structures involve complex circuit designs, which hinder their implementation.
The SC-HAD structure reduces RF chain requirements and hardware complexity by grouping antennas into subarrays that share RF links. However, the inter-subarray spacing—often several half-wavelengths—easily introduces phase ambiguity, which degrades DOA sensing accuracy. Resolving this ambiguity generally requires multiple sampling slots or complex analog phase settings, both of which severely increase system latency. For example, a common strategy involves coarse DOA estimation in the initial slot followed by ambiguity resolution through reconfigured beamformers in subsequent slots—a method unsuitable for latency-sensitive scenarios.
Although the algorithm in \cite{10164283} reduces the overall measurement latency to two sampling slots and achieves satisfactory DOA accuracy via pre-designed analog PS configurations, it still entails substantial hardware complexity and structural constraints, limiting its suitability for ultra-low-latency real-time communications. To overcome these challenges, \cite{shu2022machine} proposed a dual-layer hybrid architecture combining FD and HAD arrays to enhance ambiguity resolution under hardware constraints. Nevertheless, this method continues to impose considerable demands on hardware resources and power consumption.

To address these limitations, \cite{10767772} introduced a novel heterogeneous hybrid analog-digital (H$^2$AD) MIMO architecture. By incorporating structural heterogeneity among traditional homogeneous HAD subarrays, the H$^2$AD design achieves intrinsic phase ambiguity resolution within a single sampling slot, offering ultra-low latency. Its processing delay is comparable to FD architectures, while maintaining hardware complexity and energy consumption levels similar to conventional HAD structures. This results in a more favorable trade-off between performance and resource efficiency.
Building upon the H$^2$AD structure, \cite{10767772} further developed a "model and data-driven" learning framework for directional sensing. This framework consists of: (1) candidate DOA generation using MUSIC or DL; (2) identification of true DOAs through machine learning methods; and (3) fusion of identified DOAs for final estimation. To further enhance estimation accuracy and reduce clustering complexity, \cite{bai2024co} and \cite{10892212} proposed the co-learning-assisted multi-modal DL framework, which significantly improves both estimation performance and model generalization.




However, the weight fusion (WF) strategies employed in the aforementioned methods \cite{10767772, bai2024co, 10892212} rely on prior variance estimation or the explicit calculation of the Cramer-Rao lower bound (CRLB), which makes them less suitable for practical application scenarios where obtaining prior information is challenging. Based on the analysis and discussion above, the critical contributions of this paper can be summarized as follows:

\begin{enumerate}

    \item 
    To achieve a low-complexity and high-performance fusion of sensed DOA values from different sub-array groups in a H$^2$AD MIMO reciever with fewer or little of prior knowledge, a lightweight CRLB-ratio-WF method is proposed. Unlike the strategy in \cite{10767772}, which requires precise computation of each subarray's CRLB to determine fusion weights, the proposed method avoids the computational burden associated with real-time CRLB evaluation. Specifically, it approximates the inverse CRLB of each subarray using the reciprocal of its number of antennas, allowing for weight assignment based solely on the subarray’s structural parameters. This strategy decouples the fusion process from dependence on prior statistical knowledge (i.e., CRLB) and instead expresses the weights as a simple function of fixed array design (i.e., $M_q$), significantly reducing algorithmic complexity and reliance on prior information. Meanwhile, the proposed method maintains fusion performance and provides a scalable solution for massive array signal processing scenarios.
     
    \item A multi-branch deep neural network (MBDNN), consisting of a multi-branch fully connected neural network (MB\_FCNN) and a fusion network (fusionNet) implemented using a single linear layer, is designed for false solution elimination and true solution fusion. The MB\_FCNN comprises three input branches, each implemented as a three-layer fully connected (FC) subnetwork to perform deep nonlinear mappings on candidate angle inputs from distinct subarrays. The outputs of these branches are concatenated and passed to a shared FC regression module to produce the final DOA sensing. This design explicitly separates the network's functionality into two stages: pseudo solution elimination and true solution fusion.
    The proposed MBDNN not only preserves the unique information captured by each subarray, but also effectively exploits the cross-domain complementarity, thereby enhancing the accuracy of DOA sensing.

    \item Extensive simulations under varying SNR and snapshot conditions demonstrate that the RMSE performance of the proposed CRLB-ratio-WF method closely approaches that of the traditional CRLB-based approach, even in scenarios with limited prior information. Moreover, the proposed MBDNN method outperforms the CRLB-ratio-WF method, particularly in low signal-to-noise ratio (SNR) regions. Notably, at SNR = $-15$ dB, the estimation accuracy of the former is ten times better than the latter.
\end{enumerate}

The remainder of this paper is arranged as follows. The system model is provided in Section $\rm{\Rmnum{2}}$. Section $\rm{\Rmnum{3}}$ describes the enhanced DOA estimation framework for H$^2$AD structure. In Section $\rm{\Rmnum{4}}$, the CRLB-ratio-WF method is proposed. Multi-branch neural network-driven DOA sensing is proposed in Section $\rm{\Rmnum{5}}$. 
Finally, the experimental results and conclusions are presented in Sections $\rm{\Rmnum{6}}$ and $\rm{\Rmnum{7}}$, respectively.


\section{System model}\label{sec_sys}

\begin{figure}[!htb]
	\centering
	\includegraphics[width=3.49in]{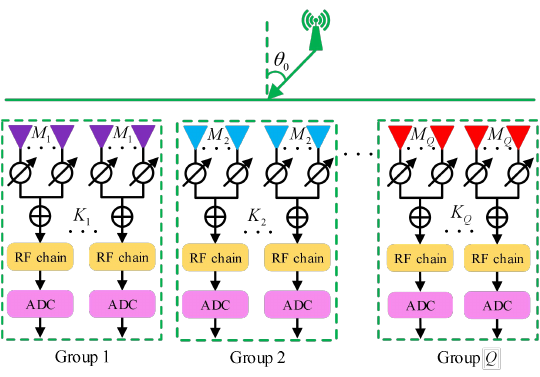}
	\caption{The H$^2$AD array has $Q$ groups and group $q$ has $K_q$ subarrays with each subarray having $M_q$ antennas, $q\in\{1,2,\cdots,Q\}$ }
    \label{system}
\end{figure}

Fig. ~\ref{system} illustrates a novel massive H\(^2\)AD MIMO architecture. Consider a far-field narrowband signal $x(t)e^{j2\pi f_ct}$ impinging on a heterogeneous antenna array, where \( x(t) \) represents the baseband signal and \( f_c\) denotes the carrier frequency.  
Assume a ULA consisting of \( N \) antennas, where all antennas are divided into \( Q \) distinct groups. Each group contains \(K_q \) subarrays and each subarray consists of \( M_q \) antennas, that is,
\begin{align}
N =\sum_{q=1}^{Q}N_q=\sum_{q=1}^{Q}K_q M_q,
\end{align}

In the H\(^2\)AD structure, \( M_1 \neq M_2 \neq \cdots \neq M_Q \). When \( M_1 = M_2 = \cdots = M_Q \), this array represents the conventional sub-connected hybrid HAD structure. Therefore, the sub-connected HAD structure is a special case of the H\(^2\)AD architecture, also referred to as a homogeneous structure, where each group in the H\(^2\)AD belongs to a homogeneous array.  The output signal of the \( q \)-th subarray is given by:

\begin{align}
	{f_{q,k}}(t) = \frac{1}{{\sqrt {{M_q}} }}\sum\limits_{m = 1}^{{M_q}} {{s_{q,k,m}}} (t){e^{ - j{\Psi _{q,k,m}}}} + {n_{q,k}}(t)
\end{align}
where $\Psi _{h,k,m}$ represents the analog beamforming phase of the \( m \)-th antenna and the \( k \)-th RF chain. $n_{q,k}(t)\sim\mathcal{C}\mathcal{N}(0,\sigma^2_w)$ is the additive white Gaussian noise (AWGN) vector. $s_{q,k,m}$ is 
the signal received by the \( k \)-th antenna of the \( m \)-th subarray in the \( q \)-th group, expressed as

\begin{equation}
    {s_{q,k,m}}(t) = x(t){e^{j2\pi {f_c}(t - {\chi _{q,k,m}})}} + {n_{q,k,m}}(t)
\end{equation}
where $\chi _{q,k,m}$ is the propagation delay influenced by the signal source's direction in relation to the array, expressed as:
\begin{align}
    {\chi _{q,k,m}} = {\chi _0} - \frac{{{d_m}\sin {\theta _0}}}{c}
\end{align}
where $c$ is the speed of light, $\chi _0$ is the propagation delay from the transmitter to a reference mark of the array.
\( d_m \) is the distance from the reference point, located at the left side of the array, to the \( m \)-th antenna of the \( k \)-th subarray, expressed as:

\begin{align}
 d_m=(km-1)d
\end{align}

The outputs \( K_q \) of all \( K_q \) subarrays are stacked and transmitted in parallel, the received vector of the \( q \) group is given by:

\begin{equation}\label{bfgtbyt}
{\mathbf{s}_{q}}({t}) = \mathbf{W}_{A,q}^H({{\bf{a}}_q}({\theta _0})x(t) + \mathbf{n}(t))
\end{equation}
where $\mathbf{W}_{A,q}$ is a block diagonal matrix, expressed as

\begin{align}
{\mathbf{W}_{A,q}} = \left[ {\begin{array}{*{20}{c}}
{{\mathbf{w}_{A,q,1}}}&0& \cdots &0\\
0&{{\mathbf{w}_{A,q,2}}}& \cdots &0\\
 \vdots & \vdots & \ddots & \vdots \\
0&0& \cdots &{{\mathbf{w}_{A,q,{K_q}}}}
\end{array}} \right] \in \mathbb{C}{^{{N_q} \times {K_q}}}
\end{align}
$\mathbf{n}(t)=\left[n_1(t), n_2(t), \ldots, n_{K_q}(t)\right]^T\in\mathbb{C}^{K_q \times 1}$ expresses an AWGN vector, $\mathbf{a}_q(\theta_0)\in\mathbb{C}^{N_q \times 1}$ is the array steering vector of $q$-th group, denoted as
\begin{align}
	\mathbf{a}_q\left(\theta_0\right)&=\left[1, e^{j \frac{2 \pi}{\lambda} d \sin \theta_0}, \cdots, e^{j \frac{2 \pi}{\lambda}(N_q-1) d \sin \theta_0}\right]^T\\
    & = {{\bf{a}}_{{I}}}\left( {{\theta _0}} \right) \otimes {{\bf{a}}_{{J}}}\left( {{\theta _0}} \right) \in \mathbb{C}{^{{N_q} \times 1}},
\end{align}
where \( \mathbf{a}_I \) is the array response vector of a uniform linear array (ULA) with \( M_q \) elements, where the antenna spacing is \( d \):
\begin{align}
{{\bf{a}}_I}\left( {{\theta _0}} \right) = {\left[ {1,{e^{j\frac{{2\pi }}{\lambda }d\sin {\theta _0}}}, \cdots ,{e^{j\frac{{2\pi }}{\lambda }({M_q} - 1)d\sin {\theta _0}}}} \right]^T}
\end{align}
\( \mathbf{a}_J \) can be represented as follows: In the \( q \)th group, each subarray can be viewed as a virtual antenna. Therefore, group \( q \) can be regarded as an array with \( K_q \) virtual antenna elements, where the interval between adjacent virtual antennas is \( M_q d = M_q \lambda / 2 \):
\begin{align}\label{cfbg}
	{{\bf{a}}_J}\left( {{\theta _0}} \right) = {\left[ {1,{e^{j\frac{{2\pi }}{\lambda }{M_q}d\sin {\theta _0}}}, \cdots ,{e^{j\frac{{2\pi }}{\lambda }({K_q} - 1){M_q}d\sin {\theta _0}}}} \right]^T}
\end{align}

And $\mathbf{w}_{A,q,k}, (k\in [1, K_q])$ is the anolog beamforming vector of $k$-th subarray, expressed as
\begin{align}
	\mathbf{w}_{A,q,k}=\frac{1}{\sqrt{M_q}}\left[e^{j\chi_{q,k,1}},e^{j\chi_{q,k,2}},\cdots,e^{j\chi_{q,k,M_q}}\right]^T,
\end{align}

Via analog-to-digital converter (ADC), the (\ref{bfgtbyt}) is rewritten as:
\begin{align}\label{frg}	\mathbf{s}_q(n)=\mathbf{W}_{A,q}^H\mathbf{a}_q(\theta_0)x(n)+\mathbf{n}(n),
\end{align}
where $n=1,2,\cdots,T$, $T$ is the number of snapshots.

\section{The Enhanced DOA Sensing Framework for H$^2$AD Structure}\label{sec_proposed}

In this section, an enhanced DOA estimator based on H$^2$AD array is proposed to rapidly eliminate phase ambiguity within a single time-slot. The true solution corresponding to each group is selected from the set of candidate solutions. The final high-performance DOA sensing is obtained by performing a weighted fusion of all true solutions.

\begin{figure}[!http]
	\centerline{\includegraphics[width=3.2in]{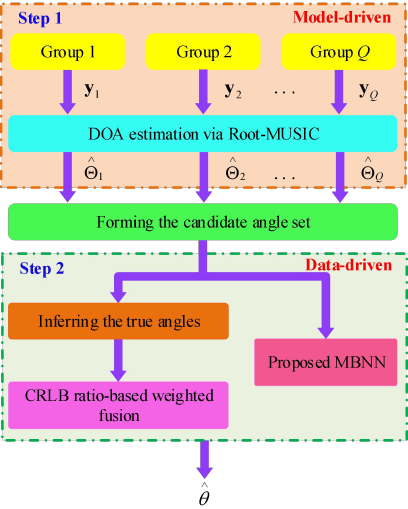}}
	\caption{The enhanced DOA estimator for H$^2$AD structure.\label{fig_alg_flow1}}
\end{figure}

As illustrated in Fig. \ref{fig_alg_flow1}, the introduced DOA sensing framework contains three major components:

1) The $Q$ subarray groups in the H\(^2\)AD structure estimate DOA value using the Root-MUSIC algorithm and yields the candidate angle sets.  

2) A MB\_FCNN is employed to remove pseudo-solutions from the candidate sets.

3) An optimal CRLB-ratio-WF method is designed based on the minimum mean squared error (MSE) to fuse the inferred \( Q \)  true solutions, obtaining an enhanced DOA estimation.

Next, we will demonstrate how phase ambiguity arises. According to the above discussion, the output signal model is 

\begin{equation}
    \mathbf{s}\left( n \right) = {\left[ {\mathbf{s}_1^T{{\left( n \right)}_{}}\;\mathbf{s}_2^T\left( t \right) \cdots \mathbf{s}_Q^T\left( n \right)} \right]^T}
\end{equation}
where $\mathbf{s}_q\left( n \right)$ is the output signal of 
$q$-th group. Since the position of the target is completely unknown, without loss of generality, all the PS are set to operate without phase adjustment, i.e., the $\mathbf{w}_{A,q,k}$ vectors of all subarrays are set as:
\begin{equation}
    \mathbf{w}_{A,q,k}=\frac{1}{{\sqrt {{M_q}} }}{\left[ {1, \cdots ,1} \right]^T}
\end{equation}
then Eq. (\ref{frg}) can be written as 

\begin{align} \label{yvfq1}
	\mathbf{s}_q(n)=\frac{1}{{\sqrt {{M_q}} }}\mathbf{a}_{M_q}(\theta_0)e_q(\theta_0)x(n)+\mathbf{n}(n),
\end{align}
where $e_q(\theta_0)$ is the gain coefficient for $q$-th subarray \cite{10767772}, defined as

\begin{align} \label{}
	e_q\left(\theta_0\right)  &=\sum_{m=1}^{M_q} e^{j \frac{2 \pi}{\lambda}(m-1) d \sin \theta_0} \\
	&=\frac{1-e^{j \frac{2 \pi}{\lambda} M_q d \sin \theta_0}}{1-e^{j \frac{2 \pi}{\lambda} d \sin \theta_0}}
\end{align}
And \( \mathbf{a}_{M_q} \) represents the steering vector of the \( h \)-th subarray in the direction \( \theta_0 \), its expression is given by Eq. (\ref{cfbg}).

\begin{figure}[!http]
	\centerline{\includegraphics[width=3.5in]{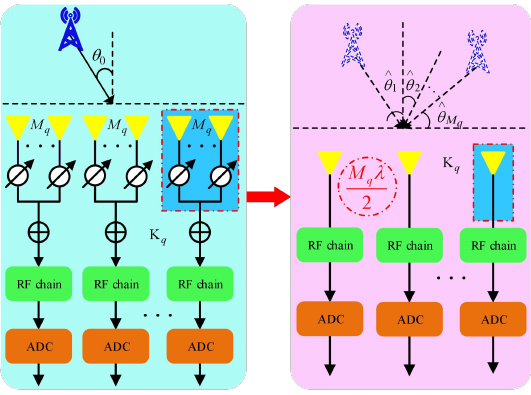}}
	\caption{The illustration of phase ambiguity formation, where each subarray of the \( q \)-th group is considered a virtual antenna, and group \( q \) is equivalent to a \( K_q \)-element ULA with an antenna spacing of \( M_q \lambda/2\).}
    \label{ambiguous}
\end{figure}

Therefore, each subarray of $q$-th group in the H\(^2\)AD array can be regarded as a virtual antenna, as shown in Fig. \ref{ambiguous}, which is equivalent to a fully digital ULA with an antenna spacing of \( M_q \) times half the wavelength. When the H\(^2\)AD receives an incident signal for DOA sensing, it will generate \( \sum\limits_{q = 1}^Q M_q \) ambiguous solutions, leading to the issue of pseudo-solution elimination.

Based on Eq. (\ref{yvfq1}), the covariance matrix of the received signal \( \mathbf{s_q(n)}\) is given by:

\begin{equation}
\begin{aligned} \label{}
	\mathbf{R}_{q} & =\mathbb{E}\left[\mathbf{s_q}(n) \mathbf{s_q}(n)^H\right] \\
	& =\frac{\sigma_x^2}{M_q} \left\|e_q\left(\theta_0\right)\right\|^2 \mathbf{a}_{M_q}\left(\theta_0\right) \mathbf{a}_{M_q}^H\left(\theta_0\right)+\sigma_v^2 \mathbf{I_v},                            
\end{aligned}
\end{equation}
where $\sigma_x^2=\mathbb{E}\left[{\left| {x\left( n \right)} \right|^2}\right]$ is the energy of the receive signal $x(n)$. In practical applications, the $\mathbf{R}_{q}$ is generally estimated using the following equation:
\begin{equation}
{{\mathop {\bf{R}}\limits^ \wedge}  _q} = \frac{1}{T}\sum\limits_{n = 1}^T {{{\mathbf{s}}_{q}}(n){{\mathbf{s}}_{q}}{{(n)}^H}} 
\end{equation}

The eigenvalue decomposition (EVD) of \( {{\mathop {\bf{R}}\limits^ \wedge}  _q} \) can still be expressed as:
\begin{align} \label{}
	{{\mathop {\bf{R}}\limits^ \wedge}  _q} =[\mathbf{U}_x \mathbf{U}_v]\Sigma[\mathbf{U}_x \mathbf{U}_v]^H,
\end{align}
where $\Sigma\in \mathbb{C}^{K_q\times K_q}$ is a diagonal matrix, designed as

\begin{equation}
\Sigma  = \left[ {\begin{array}{*{20}{c}}
{\sigma _x^2 + \sigma _v^2}&0& \cdots &0\\
0&{\sigma _v^2}& \cdots &0\\
 \vdots & \vdots & \ddots & \vdots \\
0&0& \cdots &{\sigma _v^2}
\end{array}} \right]
\end{equation}
And 
$\mathbf{U}_x \in \mathbb{C}^{{K_q} \times 1}$ and $\mathbf{U}_v\in \mathbb{C}^{{K_q} \times({K_q} - 1)}$ are the signal subspace and the noise subspace, respectively. 
Based on the relationship between the pseudo-spectrum maximum and the angle, the DOA value for each group can be estimated. For example, the MUSIC algorithm’ pseudo-spectrum \cite{schmidt1982signal} is given by:

\begin{align} \label{}
	P_{MU}(\theta)=\frac{1}{\left\|e_q(\theta)\right\|^2\left\|\mathbf{U}_v^H\mathbf{a}_{M_q}\right\|^2},
\end{align}

The peak of the pseudo-spectrum \( P_{MU} \) corresponds to the angle \( \theta \), which is the estimated DOA value of the $q$-th group. However, the linear search process of MUSIC causes higher computational complexity, while the Root-MUSIC \cite{1993The} achieves fast and search-free DOA estimation by finding the roots of a polynomial.
Therefore, in this paper, Root-MUSIC is used to obtain the feasible solution set.

Let $\mathbf{F}=\mathbf{U}_v\mathbf{U}_v^H$, the root-finding expression can be written as:

\begin{equation}\label{p}
\begin{aligned} 	y(\theta)&=\left\|e_q(\theta)\right\|^2(\theta)\mathbf{a}_{M_q}^H(\theta)\mathbf{U}_v\mathbf{U}_v^H\mathbf{a}_{M_q}(\theta)\\
	&=\frac{2-e^{-j\frac{2 \pi}{\lambda} M_q d \sin \theta_0}-e^{j\frac{2 \pi}{\lambda} M_q d \sin \theta_0}}{2-e^{-j\frac{2 \pi}{\lambda} d \sin \theta_0}-e^{j\frac{2 \pi}{\lambda} d \sin \theta_0}}\\
    &\cdot\sum_{i=1}^{K_q}\sum_{j=1}^{K_q}z^{-(i-1)}\mathbf{F}_{ij}z^{j-1}
\end{aligned}
\end{equation}

Define $z_q=e^{j\frac{2 \pi}{\lambda} M_q d \sin \theta_0}$, the above polynomial is written as:

\begin{equation}
\begin{aligned} 	
y(z_q)
	&=\frac{2-z_q^{-1}-z_q}{2-z_q^{-\frac{1}{M_q}}-z_q^{\frac{1}{M_q}}}\sum_{i=1}^{K_q}\sum_{j=1}^{K_q}z_q^{-(i-1)}\mathbf{F}_{ij}z_q^{j-1}\\
   &\triangleq y(\varphi_q)=0,
\end{aligned}
\end{equation}
where $\varphi_q$ is denoted as 

\begin{align} \label{}
	\varphi_q=\frac{2 \pi}{\lambda} M_q d \sin \theta,
\end{align}

The Eq. (\ref{p}) consists of $2(K_q-1)$ roots, i.e., $\left\{z_1, z_2, \cdots, z_{2K_q-2}\right\}$. Then, the corresponding emitter angle set can be obtained:

\begin{align} \label{}
	{\mathbf{\hat \Theta }}_{MU}=\left\{\hat{\theta}_1, \hat{\theta}_2, \cdots, \hat{\theta}_{2K_q-2}\right\},
\end{align}
where
\begin{align} \label{cfhbytjn}
	\hat{\theta}_i=\arcsin \left(\frac{\lambda \arg z_i}{2 \pi M_q d}\right)
\end{align}

According to the literature \cite{shuHADDOA2018tcom}, \( y(\varphi) \) is a periodic function with a period of \( 2\pi \), i.e., \( y(\hat{\varphi}_q) = y(\hat{\varphi}_q + 2\pi j) \). Additionally, each subarray in the \( q \)-th group is considered a virtual antenna, and group \( q \) is equivalent to a \( K_q \)-element ULA with an antenna spacing of \( M_q \lambda/2 \), resulting in \( M_q \) candidate solutions. To handle this phase ambiguity, Eq. (\ref{cfhbytjn}) is transformed into:
\begin{align} \label{}
	\hat{\theta}_{q, j_q}=\arcsin \left(\frac{\lambda\left(\arg (e^{j\hat{\varphi}_q})+2 \pi j_q\right)}{2 \pi M_q d}\right) .
\end{align}
where $j_q\in\{1,2,\cdots,M_q\}$ is the phase ambiguity factor.

Then, the feasible solutions of group $q$ can be expressed as:
\begin{equation}
    \begin{aligned}
    {{\mathbf{\hat \Theta }}_q} &= \left\{ {{{\left. {{{\hat \theta }_{q,{j_q}}}} \right|}_{{j_q} \in \left[ {1,{M_q}} \right]}}} \right\} \in  \mathbb{R}{^{{M_q}}}\\
 &= \left\{ {{{\hat \theta }_{q,1}},{{\hat \theta }_{q,2}}, \cdots ,{{\hat \theta }_{q,{M_q}}}} \right\}
\end{aligned} 
\end{equation}

The whole $Q$ subarray groups are combined can obtain:
\begin{align}\label{angleCandAvfll}
	\begin{split}
		\left \{
		\begin{array}{ll}
			\frac{2\pi}{\lambda}M_1d\sin{\hat \theta }_{1,j_1}=\hat{\varphi}_1+2\pi j_1\\
			\frac{2\pi}{\lambda}M_2d\sin{\hat \theta }_{2,j_2}=\hat{\varphi}_2+2\pi j_2\\
			\quad\quad\quad\vdots\quad\quad\quad\quad\quad\quad\quad\vdots\\
			\frac{2\pi}{\lambda}M_Qd\sin{\hat \theta }_{Q,j_Q}=\hat{\varphi}_Q+2\pi j_Q
		\end{array}
		\right.
	\end{split}
\end{align}
where \( \hat{\theta}_{q, j_q} \) is the \( j_q \)-th candidate angle of the \( q \)-th subarray. 

Based on this, the entire set of candidate solutions is given by:
\begin{align} \label{}
	\mathbf{\hat \Theta }=\left\{\mathbf{\hat \Theta }_{1}, \mathbf{\hat \Theta }_{2} \cdots, \mathbf{\hat \Theta }_{Q}\right\}
\end{align}

Then, the Eq. (\ref{angleCandAvfll}) is transformed into:

\begin{align}
	\left\{ {\begin{array}{*{20}{l}}
{{M_1} \cdot \left( {\frac{{2\pi }}{\lambda }d\sin {{\hat \theta } _{1,{j_1}}}} \right) \equiv {{\hat \varphi }_1}{{_{}}_{}}{{_{}}^{}}\left( {{{\bmod }_{}}2\pi } \right)}\\
{{M_2} \cdot \left( {\frac{{2\pi }}{\lambda }d\sin {{\hat \theta } _{2,{j_2}}}} \right) \equiv {{\hat \varphi }_2}^{}\left( {{{\bmod }_{}}2\pi } \right)}\\
{\quad \quad \quad  \vdots \quad \quad \quad \quad \quad \quad \quad  \vdots }\\
{{M_Q} \cdot \left( {\frac{{2\pi }}{\lambda }d\sin {{\hat \theta } _{Q,{j_Q}}}} \right) \equiv {{\hat \varphi }_Q}^{}\left( {{{\bmod }_{}}2\pi } \right)}
\end{array}} \right.
\end{align}

Let $x = \frac{{2\pi }}{\lambda }d\sin {\theta _{q,{j_q}}}$, a system of linear congruences in \( x \) can be obtained:

\begin{equation}\label{cdvggt}
    \left\{ \begin{array}{l}
x \equiv {\frac{{{{\hat \varphi }_1}}}{{{M_1}}}^{}}\left( {{{\bmod }_{}}\frac{{2\pi }}{{{M_1}}}} \right)\\
x \equiv {\frac{{{{\hat \varphi }_2}}}{{{M_2}}}^{}}\left( {{{\bmod }_{}}\frac{{2\pi }}{{{M_2}}}} \right)\\
 \vdots \\
x \equiv {\frac{{{{\hat \varphi }_Q}}}{{{M_Q}}}^{}}\left( {{{\bmod }_{}}\frac{{2\pi }}{{{M_Q}}}} \right)
\end{array} \right.
\end{equation}

According to the Chinese Remainder Theorem (CRT), a necessary and sufficient condition for Eq. (\ref{cdvggt}) to have a unique true solution is that \(\frac{{2\pi }}{{{M_q}}}\) are co-prime, i.e.,

\begin{equation}
    {\rm{GCD}}\left( {\frac{{2\pi }}{{{M_q}}},\frac{{2\pi }}{{{M_k}}}} \right) = 1,\quad \forall q \ne k
\end{equation}
where $\rm{GCD}$ is the greatest common divisor. Since \( 2\pi \) is a constant, the coprimality of \( \frac{{2\pi }}{{M_q}} \) is equivalent to the coprimality of \( M_q \), expressed as: 

\begin{equation}
    \mathrm{GCD}\left( {{M_q},{M_k}} \right) = 1, \quad \forall q \ne k 
\end{equation}
Thus, to ensure that all true solutions can be clustered into a single class while the remaining false solutions form other classes, all \( M_q \) should be prime numbers, or the greatest common divisor (GCD) of all \( M_q \) should be equal to $1$.

Given the preceding analysis, the entire candidate angle set \( {\mathbf{\hat \Theta }} \) consists of \( \sum\limits_{q = 1}^Q M_q \) solutions. Every candidate set contains \( M_q - 1 \) pseudo-solutions and one true estimated angle.  
Therefore, the primary task of this research is to identify the true angle from the candidate angle set.
For the $q$-th group, the true solution and pseudo angles can be expressed by

\begin{align} \label{qqqqq}
	\hat{\theta}_{t,q}=\theta_0+\eta_q
\end{align}	
\begin{align} \label{cc}
\hat{\theta}_{p,q,m}=\theta_0+\eta_q+\Xi _{q,m},~m=1,2,\cdots,M_q-1,
\end{align}
where $\eta_q$ is the measurement error and $\Xi _{q,m}$ is the offset of the pseudo-solution determined by \( M_q \) and \( m \).
For $q$-th pseudo-solution $\hat{\theta}_{p,q,m}$, when $j_q=m$, $\sin\hat{\theta}_{p,q,m}$ is: 
\begin{equation}\label{efedefe}
    \sin{\hat \theta _{p,q,m}} = \frac{\lambda }{{2\pi {M_q}d}}\left( {{{\hat \varphi }_q} + 2\pi m} \right)
\end{equation}

Assuming the absence of noise, i.e., $\eta_q\rightarrow 0$, then ${{\hat \theta }_{p,q,m}}= {{\theta _0} + {\Xi  _{q,m}}} $.
Based on the Taylor expansion, \(\sin {{\hat \theta }_{p,q,m}}\) can 
be approximated as:

\begin{equation}\label{defe}
   \sin \left( {{\theta _0} + {\Xi  _{q,m}}} \right) \approx \sin {\theta _0} + \cos {\theta _0}\sin {\Xi  _{q,m}}
\end{equation}

Substituting Eq. (\ref{defe}) and $\sin {\theta _0} = \frac{\lambda }{{2\pi {M_q}d}}{{\hat \varphi }_q}$ into Eq. (\ref{efedefe}) yields:



\begin{equation}
    {\Xi  _{q,m}} \approx \frac{\lambda }{d} \cdot \frac{m}{{{M_q}}},~m \in [1,M_q-1]
\end{equation}

Besides, due to $\mathrm{GCD}\left( {{M_q},{M_k}} \right) = 1$ $(\forall q \ne k)$, the equation $m M_k=n M_q$ has no solution, i.e.,

\begin{equation}
   {\frac{m}{{{M_q}}} \ne \frac{n}{{{M_k}}}}
\end{equation}
where $m\in[1,M_q-1]$, $n\in[1,M_k-1]$. Thus, the offset of pseudo-solution \( \Xi_{q,m} \) and \( \Xi_{k,n} \) will not overlap.

In summary, as \(\eta_q \rightarrow 0\), the true solution $\{\hat{\theta}_{t,q}, q\in [1,Q]\}$ for all subarrays must converge to the same value \(\theta_0\). Meanwhile, to ensure that the pseudo-solutions, $\hat{\theta}_{p,q,m}=\theta_0+\Xi _{q,m}$, are distinguishable from the true solution, it is required that \({\Xi  _{q,m}} \ne 0\) and \({\Xi  _{q,m}} \ne {\Xi  _{k,n}}\) for \(\forall\left( {q,m} \right) \ne \left( {k,n} \right)\). Hence, the true angle formed is given as:

\begin{align} \label{bb}
{\hat{\theta}_{t,1}} \approx {\hat{\theta}_{t,2}} \approx  \cdots  \approx {\hat{\theta}_{t,Q}} \approx {\theta _0}
\end{align}
and pseudo-solutions can be repressed as:
\begin{align} 
{\hat{\theta}_{p,1,m}} \ne {\hat{\theta}_{p,2,m}} \ne  \cdots  \ne {\hat{\theta}_{p,Q,m}} \ne {\theta _0}
\end{align}   
This implies that the error between the true angle approaches zero, allowing the use of a similarity measure or distance metric to estimate the true emitter angle.

Thus, the true angle set can be specified as
\begin{equation}\label{eeeee}
	\mathbf{\hat \Theta }_{t}=\left\{ \hat{\theta}_{t,1},\hat{\theta}_{t,2},\cdots,\hat{\theta}_{t,Q} \right\}
\end{equation}
where $\hat{\theta}_{t,q}$ denotes the selected true emitter direction of group $q$.

\section{Proposed CRLB-Ratio-WF Method}
In this section, to achieve more efficient and low-complexity DOA measurement with fewer or little prior information, a CRLB-ratio-WF approach is proposed to combine all the true solutions.

Firstly, the inferred $Q$ true angles are fused to achieve the final DOA measurement, expressed as:

\begin{align} \label{thetaEstConbine}
	\hat{\theta}=\sum_{q=1}^{Q} w_{q}\hat{\theta}_{t,q} 
\end{align}

The MSE of the $\hat{\theta}$ is
\begin{equation} 
\begin{aligned} \label{MAEA}
	\mathbf{MSE}(\hat{\theta})&=\mathbb{E}\left[\left(\hat{\theta}-\theta_0\right)^2\right]=\sum_{q=1}^{Q} w_{q}^2
 \mathbb{E}\left[\left(\hat{\theta}_{t,q}-\theta_0\right)^2\right]\\
 &=\sum_{q=1}^{Q} w_{q}^2\mathbf{MSE}(\hat{\theta}_{t,q})\ge \sum_{q=1}^{Q} w_{q}^2CRLB_{q}.
\end{aligned}
\end{equation}
Thus, the issue is transformed into
\begin{align} \label{w_op}
	&\min_{w_q}~~~\sum_{q=1}^{Q} w_{q}^2CRLB_{q}\nonumber\\
	&~s.t. ~~~~\sum_{q=1}^{Q} w_{q}=1,
\end{align}
where
\begin{align} \label{evwqFinal2}
{w_q} = \frac{{CRLB_q^{ - 1}}}{{\sum\limits_{q = 1}^Q C RLB_q^{ - 1}}},
\end{align}

Then, we mainly concentrate on the derivation of Eq. (\ref{evwqFinal2}).

Let
\begin{equation}\label{nxueb}
   \frac{CRLB_q}{CRLB_1} =r_q 
\end{equation}

Thus,
Eq. (\ref{evwqFinal2}) can be converted into
\begin{align}\label{vfvnxueb} 
{w_q} = \frac{{r_q^{ - 1}CRLB_1^{ - 1}}}{{\sum\limits_{q = 1}^Q {r_q^{ - 1}C} RLB_1^{ - 1}}} = \frac{{r_q^{ - 1}}}{{\sum\limits_{q = 1}^Q {r_q^{ - 1}} }},
\end{align}

According to \cite{10767772}, $CRLB_q$ and $CRLB_1$ are respectively expressed via Eq. (\ref{crlb}) and Eq. (\ref{crlb1})

\begin{figure*}
\begin{equation}\label{crlb}
CRL{B_q} = \frac{{{\lambda ^2}{M_q}{\mathbf{\Upsilon}_q}}}{{8L{\pi ^2}{\bf{SNR}}{{\cos }^2}{\theta _0}\left[ {\frac{{{{\left\| {{e_q}({\theta _0})} \right\|}^4}M_q^2K_q^2\left( {K_q^2 - 1} \right){d^2}}}{{12}} + \frac{{{M_q}{K_q}}}{{{\mathbf{\Upsilon} _q}}}\left( {{{\left\| {{e_q}({\theta _0})\vartheta } \right\|}^2} + {K_q}\Re \left\{ {e_q^2({\theta _0})\vartheta } \right\}} \right)} \right]}}
\end{equation}
\end{figure*}

\begin{figure*}
\begin{equation}\label{crlb1}
CRL{B_1} = \frac{{{\lambda ^2}{M_1}{\mathbf{\Upsilon} _1}}}{{8L{\pi ^2}{\bf{SNR}}{{\cos }^2}{\theta _0}\left[ {\frac{{{{\left\| {{e_1}({\theta _0})} \right\|}^4}M_1^2K_1^2\left( {K_1^2 - 1} \right){d^2}}}{{12}} + \frac{{{M_1}{K_1}}}{{{\mathbf{\Upsilon} _1}}}\left( {{{\left\| {{e_1}({\theta _0})\vartheta } \right\|}^2} + {K_1}\Re \left\{ {e_1^2({\theta _0})\vartheta } \right\}} \right)} \right]}}
\end{equation}
\end{figure*}
where 
\begin{align}
	\vartheta = \sum_{m=1}^{M_q}\left(m-1\right)d e^{-j\frac{2\pi}{\lambda}(m-1)d\sin{\theta _0}}
\end{align}
\begin{align}
	\mathbf{\Upsilon}_q  = {M_q} + {K_q}{M_q}{\left\| {{e_q}({\theta _0})} \right\|^2}
\end{align}

\begin{align}
	\mathbf{\Upsilon}_1  = {M_1} + {K_1}{M_q}{\left\| {{e_1}({\theta _0})} \right\|^2}
\end{align}


Using Euler's formulas 
and the Taylor approximation, ${\left\| {{e_q}\left( {{\theta _0}} \right)} \right\|^2}$ can be simplified to:



\begin{equation}
  {\left\| {{e_q}\left( {{\theta _0}} \right)} \right\|^2} \approx \frac{{{{\left( {\frac{\pi }{\lambda }{M_q}d\sin {\theta _0}} \right)}^2}}}{{{{\left( {\frac{\pi }{\lambda }d\sin {\theta _0}} \right)}^2}}} = M_q^2
\end{equation}

Similarly, 
\begin{equation} 
 {\left\| {{e_1}\left( {{\theta _0}} \right)} \right\|^2} \approx \frac{{{{\left( {\frac{\pi }{\lambda }{M_1}d\sin {\theta _0}} \right)}^2}}}{{{{\left( {\frac{\pi }{\lambda }d\sin {\theta _0}} \right)}^2}}} = M_1^2
\end{equation}

Thus

\begin{equation}\label{werwrw}
    \mathbf{\Upsilon}_q \approx M_q + {K_q}M_q^3
\end{equation}

\begin{equation}\label{w2erwrw}
    \mathbf{\Upsilon}_1 \approx M_1 + {K_1}M_1^3
\end{equation}

For H$^2$AD array in this research, $K_1 = K_2=\cdots = K_Q$, thus Eq. (\ref{crlb}) and Eq. (\ref{crlb1}) is converted into Eq. (\ref{crlbue7}) and Eq. (\ref{crlbue8}), respectively: 
\begin{figure*}
\begin{equation}\label{crlbue7}
CRL{B_q} \approx \frac{{{\lambda ^2}{M_q}(M_q + {K}M_q^3)}}{{8L{\pi ^2}{\bf{SNR}}{{\cos }^2}{\theta _0}\left[ {\frac{{{{\left\| {{e_q}({\theta _0})} \right\|}^4}M_q^2K^2\left( {K^2 - 1} \right){d^2}}}{{12}} + \frac{{{M_q}{K}}}{{{M_q + {K}M_q^3}}}\left( {{{\left\| {{e_q}({\theta _0})\vartheta } \right\|}^2} + {K}\Re \left\{ {e_q^2({\theta _0})\vartheta } \right\}} \right)} \right]}}
\end{equation}
\end{figure*}

\begin{figure*}
\begin{equation}\label{crlbue8}
CRL{B_1} \approx \frac{{{\lambda ^2}{{M_1}(M_1 + {K}M_1^3)}}}{{8L{\pi ^2}{\bf{SNR}}{{\cos }^2}{\theta _0}\left[ {\frac{{{{\left\| {{e_1}({\theta _0})} \right\|}^4}M_1^2K^2\left( {K^2 - 1} \right){d^2}}}{{12}} + \frac{{{M_1}{K}}}{{{M_1 + {K}M_1^3}}}\left( {{{\left\| {{e_1}({\theta _0})\vartheta } \right\|}^2} + {K}\Re \left\{ {e_1^2({\theta _0})\vartheta } \right\}} \right)} \right]}}
\end{equation}
\end{figure*}

For the $CRLB_q$ and $CRLB_1$, the denominator (proportional to $M_q$), called $D_q$ and $D_1$, can be approximated as:


\begin{equation}
\begin{aligned}
    D_q &= \frac{\| e_q(\theta_0) \|^4 M_q^2 K^2 (K^2 - 1) d^2}{12}\\ &\approx \frac{ M_q^6 K^2 (K^2 - 1) d^2}{12}
\end{aligned}
\end{equation}

\begin{equation}
\begin{aligned}
    D_1 &= \frac{\| e_1(\theta_0) \|^4 M_1^2 K^2 (K^2 - 1) d^2}{12}\\ &\approx \frac{ M_1^6 K^2 (K^2 - 1) d^2}{12}
\end{aligned}
\end{equation}

Therefore, the expression for $CRLB_q$ and $CRLB_1$ can be further simplified as:

\begin{equation}
\begin{aligned}
    CRLB_q &\approx \frac{\lambda^2 M_q (M_q + {K}M_q^3)}{8 L \pi^2 \mathbf{SNR} \cos^2 \theta_0 }\cdot \frac{1}{D_q}
   \\ &\approx \frac{{12{\lambda ^2}K^2}}{{8L{\pi ^2}{\bf{SNR}}{{\cos }^2}{\theta _0}{K^2}({K^2} - 1){d^2}}} \cdot \frac{{{M_q}{{(M_q^3 )}}}}{{M_q^6}}
\end{aligned}
\end{equation}

\begin{equation}
\begin{aligned}
    CRLB_1 &\approx \frac{\lambda^2 M_1 (M_1 + {K}M_1^3)}{8 L \pi^2 \mathbf{SNR} \cos^2 \theta_0 }\cdot \frac{1}{D_q}
   \\ &\approx \frac{{12{\lambda ^2}K^2}}{{8L{\pi ^2}{\bf{SNR}}{{\cos }^2}{\theta _0}{K^2}({K^2} - 1){d^2}}} \cdot \frac{{{M_1}{{(M_1^3 )}}}}{{M_1^6}}
\end{aligned}
\end{equation}

Based on the above analysis, Eq. (\ref{nxueb}) can be approximated as 

\begin{equation}\label{defrfgrgr}
   r_q=\frac{CRLB_q}{CRLB_1} \approx \frac{{M_1^2}}{{M_q^2}}
\end{equation}

And the Eq. (\ref{vfvnxueb}) can be further transformed into:

\begin{equation} \label{vfbgnbhn}
{{\hat w}_q} \approx \frac{{{{\left( {\frac{{M_1^2}}{{M_q^2}}} \right)}^{ - 1}}}}{{\sum\limits_{q = 1}^Q {{{\left( {\frac{{M_1^2}}{{M_q^2}}} \right)}^{ - 1}}} }} = \frac{{M_q^2}}{{\sum\limits_{q = 1}^Q {M_q^2} }},
\end{equation}

Therefore, Eq. (\ref{thetaEstConbine}) is rewritten as a weighted fusion method based on the CRLB ratio:

\begin{align} \label{hnh}
	\hat{\theta}=\sum_{q=1}^{Q} \hat w_{q}\hat{\theta}_{t,q} 
\end{align}

It is obvious that this method does not rely on extensive prior knowledge (such as SNR) or pre-calculated variances, and only requires knowledge of the number of antennas in each subarray. As a result, it requires less information and has lower computational complexity, resulting in more efficient and cost-effective DOA sensing.

According to the above-mentioned analysis, the proposed enhanced DOA estimation method has three stages: 1) generate the feasible angle sets. 2) select the true emitter direction for each group via the available clustering method. 3) merge all true angles by CRLB-ratio-WF. The entire algorithm is summarized in Algorithm \ref{alg:GS}.

\begin{algorithm}[t]
	\caption{Proposed CRLB-ratio-WF for DOA sensing.}\label{alg:GS}
	\begin{algorithmic}
		\STATE 
		\STATE {\textbf{Input:}}$~\mathbf{s}(n),~ n=1,2,\cdots,T.$
		\STATE \hspace{0.5cm} \textbf{Initialization:} ~split $\mathbf{s}(n)$ into $\mathbf{s}_{q}(n)$, $q=1,2,\cdots,Q$.
		\STATE \hspace{0.5cm} \textbf{for} $q=1,2,\cdots,Q$ \textbf{do},
		\STATE \hspace{1cm} apply the root-MUSIC algorithm to $\mathbf{s}_q(n)$ to get the feasible angle set $\hat{\Theta}_q$.
		\STATE \hspace{0.5cm} \textbf{end for}
            \STATE \hspace{0.5cm} 
            Use the global minimum distance clustering in \cite{10767772} to acquire the true solution class $\hat{\Theta}_{t}'$.
            \STATE \hspace{0.5cm} 
            The CRLB-ratio-WF coefficient ${\hat w_q}$ is approximated via Eq. (\ref{crlb}) to Eq. (\ref{vfbgnbhn}).
            \STATE \hspace{0.5cm} 
            Substitute Eq. (\ref{vfbgnbhn}) and $\hat{\Theta}_{t}'$ into Eq. (\ref{hnh}) to achieve $\hat{\theta}$.
		\STATE {\textbf{Output:}} $\hat{\theta}$
	\end{algorithmic}
\end{algorithm}

\section{Proposed MBDNN-Driven DOA Sensing}\label{sec_CNN}
In this section, a MBDNN structure, consisting of a MB\_FCNN and a single linear layer-based fusionNet, is designed with integrated functionalities of false solution elimination and true solution fusion, aiming to achieve higher DOA sensing accuracy. 

\subsection{Datasets}

To train the proposed MBDNN-based model, we constructed a dataset comprising DOA candidate estimation obtained from H$^2$AD array structure composed of $Q$ subarray groups, under various SNR conditions and incident angle scenarios. The dataset generation process is as follows:

First, similar to other DL-based DOA sensing methods, we discretize the target DOA values with a step size of $s=1^\circ$. Assuming the angle search range is defined as $[\theta_{\rm{min}},\theta_{\rm{max}}]$, then:
\begin{equation}
    \rm{N}_\theta=\frac{{{\theta _{\rm{max}}} - {\theta _{\rm{min}}}}}{s}
\end{equation}

At each angle $\theta$, the incident narrowband ULA signal is divided into $Q$ subarray groups, and the corresponding received signal vectors are formed. For each group of received signals, the Root-MUSIC algorithm is applied to extract candidate DOA estimates. Considering the presence of phase ambiguity among subarrays, the extracted angle estimation serve as ambiguous candidate DOAs. For example, for the received signal $\mathbf{s}_q(n)$ from $q$-th subarray group, the Root-MUSIC algorithm is employed to generate
$M_q$ feasible angle estimates, denoted as $\left\{ {{{\hat \theta }_{q,1}},{{\hat \theta }_{q,2}}, \cdots ,{{\hat \theta }_{q,{M_q}}}} \right\}$. The feature vector of each data sample is formed by concatenating all the candidate angles estimated via Root-MUSIC from $Q$ subarray groups. Therefore, all the 
$Q$ groups generate $\sum\limits_{q = 1}^Q {{M_q}}$
values $\mathbf{\hat \Theta }_d=\left\{\mathbf{\hat \Theta }_{d,1}, \mathbf{\hat \Theta }_{d,2} \cdots, \mathbf{\hat \Theta }_{d,Q}\right\}$, where $d$ indicates the $d$-th training sample and $\mathbf{\hat \Theta }_{d,q}=\left\{ {{{\hat \theta }_{d,q,1}},{{\hat \theta }_{d,q,2}}, \cdots ,{{\hat \theta }_{d,q,{M_q}}}} \right\}$. The corresponding label is the truth DOA simulated for that sample.
To enhance the generalization capability of the model, the number of Monte Carlo experiment is set to $T$. In addition, considering the performance of the MBDNN under varying SNR conditions, the network is trained over a defined SNR range. Assuming $\rm{SNR}\in[\rm{SNR}_{min},\rm{SNR}_{max}]$, with a sampling interval of $\nabla \rm{SNR}$, then:

\begin{equation}
    \rm{N}_{\rm{SNR}}=\frac{{{\rm{SNR} _{max}} - {\rm{SNR} _{min}}}}{\nabla \rm{SNR}}
\end{equation}

For each SNR level and each truth angle, $T$ independent simulations are conducted. Consequently, the total size of the generated training dataset is:
\begin{equation}
    \rm{N}_{\rm{total}} =\mathit{T} \rm{N}_\theta \rm{N}_{\rm{SNR}}
\end{equation}

According to the above description, the input and output of the MB\_FCNN are the complete set of candidate angles $\mathbf{\hat \Theta }=\left\{ {{{\mathbf{\hat \Theta } }_d}} \right\}_{d = 1}^{{{\rm{N}}_{{\rm{total}}}}}$ and the expected true solution for each group $\mathbf{\hat \Theta }_{t}=\left\{ {{{\mathbf{\hat \Theta } }_{d,t}}} \right\}_{d = 1}^{{{\rm{N}}_{{\rm{total}}}}}$, where  
$\mathbf{\hat \Theta }_{d,t}=\left[ \hat{\theta}_{d,t,1},\hat{\theta}_{d,t,2},\cdots,\hat{\theta}_{d,t,Q} \right]^T$, respectively. Therefore, the entire training dataset for the MB\_FCNN is:

\begin{equation}
    \mathbb{T}{_{{\rm{MB\_FCNN}}}} = \left\{ {\left. {\left( {{{\mathbf{\hat \Theta } }_d}, {{\mathbf{\hat \Theta } }_{d,t}}} \right)} \right|d = 1,2, \cdots ,{{\rm{N}}_{{\rm{total}}}}} \right\}
\end{equation}
where ${\left( {{{\mathbf{\hat \Theta } }_d}, {{\mathbf{\hat \Theta }}_{d,t}}} \right)}$ denotes the $d$-th training data and label pair.

Similarly, the training dataset for FusionNet is represented as:
\begin{equation}
   \mathbb{T}{_{{\rm{FusionNet}}}} = \left\{ {\left. {\left( {{{\mathbf{\hat \Theta } }_{d,t}},\theta_d } \right)} \right|\theta_d  \in \left[ {{\theta _{\rm{min}}},{\theta _{\rm{max}}}} \right]} \right\}_{d = 1}^{{{\rm{N}}_{{\rm{total}}}}}
\end{equation}

\subsection{Network Architecture Design}

According to the phase ambiguity generation process illustrated in Fig. \ref{ambiguous}, the candidate set ${\mathbf{\hat \Theta }}_d$ contains ${\sum\limits_{q = 1}^Q {{M_q}} } - Q$ pseudo-solutions. Therefore, the key to eliminating the pseudo-solutions lies in accurately inferring the $Q$ true angles from the $\sum\limits_{q = 1}^Q {{M_q}}$ candidate angles. Given the inherent nonlinear relationship involved in this problem, it can be modeled as:

\begin{equation}
	{{\mathbf{\hat \Theta } }_{d,t}}=f_{ep}(\mathbf{\hat \Theta })\label{nonlinear mapping}
\end{equation}
where $f_{ep}\left(  \cdot  \right): \sum\limits_{q = 1}^Q {{M_q}} \to Q$ is an unknown nonlinear mapping function. Due to the inability to explicitly model the underlying mapping mechanism between $\mathbf{\hat \Theta }$ and ${{\mathbf{\hat \Theta } }_{d,t}}$, conventional linear optimization methods are often inapplicable. To address such nonlinear mapping problems, neural networks \cite{bishop1995neural, 9457195,rumelhart1986learning} have demonstrated exceptional capabilities in efficiently learning complex relationships by its multi-layer nonlinear transformation. Based on this, a MB\_FCNN is introduced to learn the nonlinear mapping defined in Eq. (\ref{nonlinear mapping}).

\begin{figure}[!http]
	\centerline{\includegraphics[width=3.5in]{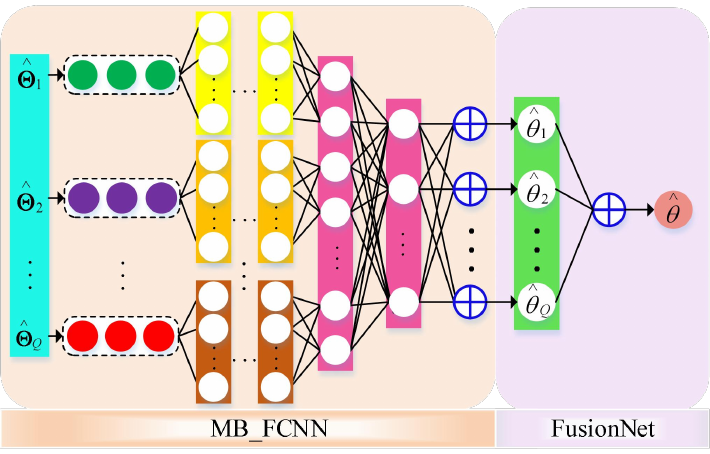}}
	\caption{The proposed MBDNN for DOA sensing,where the MB\_FCNN is used to eliminate pseudo-solutions, and FusionNet performs true solution fusion.\label{fig_alg_flow5}}
\end{figure}

As shown in Fig. \ref{fig_alg_flow5}, the MB\_FCNN is designed to infer the true DOA within each group from multiple candidate solutions, while FusionNet aggregates the inferred true angles from all groups. This framework offers an efficient and intelligent strategy for filtering and fusing candidate solutions across multiple groups. 

Firstly, to accommodate the structure of the MB\_FCNN, the overall input feature vector ${\mathbf{\hat \Theta }}_d \in {\mathbb{R}^{\sum\limits_{q = 1}^Q {{M_q}} }}$ is partitioned into $Q$ sub-vectors according to the number of subarray groups $Q$:

\begin{equation}
  {\mathbf{x}_q} = {\mathbf{\hat \Theta }}_d\left[ {\sum\limits_{i = 1}^{q - 1} {{M_i}} :\sum\limits_{i = 1}^q {{M_i}} } \right],q = 1,2, \ldots ,Q
\end{equation}
where ${\mathbf{\hat \Theta }}_d\left[ a:b \right]$ denotes the feature vector spanning from the $a$-th to the $(b-1)$-th dimension. Each sub-vector $\mathbf{x}_q$ is treated as an independent input branch of the MB\_FCNN and fed into the corresponding sub-network, enabling the extraction of local features from the candidate DOA estimation of each subarray group. In other words, the total input layer of the MB\_FCNN consists of $\sum\limits_{q = 1}^Q {{M_q}}$ neurons, with the input of each individual sub-network containing $M_1, M_2, \ldots, M_Q$ neurons, respectively. And train all sub-networks in parallel. The output layer of the MB\_FCNN comprises $Q$ neurons, which are used to predict the true angle vector ${{\mathbf{\hat \Theta }}_{d,t}}$. Thus, Eq. (\ref{nonlinear mapping}) is rewritten
as

\begin{equation}
	{{\mathbf{\hat \Theta } }_{d,t}}=f_{\rm{MF}}(\mathbf{x}_1,\mathbf{x}_2,\ldots,\mathbf{x}_Q;\mathbb{W})
\end{equation}
where $\mathbb{W}$ includes all weight and bias parameters.
More specifically, each subarray corresponds to an independent input branch, with its input consisting of the $M_q$ candidate angles estimated by $q$-th subarray. Each branch employs a three-layer FC structure to perform nonlinear transformations. The feature mapping of the $q$-th subarray branch is denoted as:

\begin{equation}
{{\bf{h}}_q} = \sigma \left( {{\bf{W}}_q^{(3)} \sigma \left( {{\bf{W}}_q^{(2)}  \sigma \left( {{\bf{W}}_q^{(1)} {{\bf{x}}_q} + {\bf{b}}_q^{(1)}} \right) + {\bf{b}}_q^{(2)}} \right) + {\bf{b}}_q^{(3)}} \right)
\end{equation}
where $\sigma$ denotes the ReLU activation function, defined as ${\rm{ReLU}}(\upsilon) = \max(0, \upsilon)$. The input $\mathbf{x}_q \in \mathbb{R}^{M_q}$ represents the candidate angles from the $q$-th subarray. The weight $\mathbf{W}_q^{(1)} \in \mathbb{R}^{M_q \times 4M_q}$,  $\mathbf{W}_q^{(2)} \in \mathbb{R}^{4M_q \times 2M_1}$ and 
$\mathbf{W}_q^{(3)} \in \mathbb{R}^{2M_1 \times M_q}$correspond to the first and second fully connected layers, respectively. The bias vectors $\mathbf{b}_q^{(1)} \in \mathbb{R}^{4M_q}$, $\mathbf{b}_q^{(2)} \in \mathbb{R}^{2M_1}$ and $\mathbf{b}_q^{(3)} \in \mathbb{R}^{M_q}$ are the associated bias parameters.  
This three-layer FC structure adopts an "expansion–compression" strategy (e.g., $M_1 \rightarrow 4M_1 \rightarrow 2M_1\rightarrow M_1$) to enhance the network’s nonlinear feature representation capacity.

Then, the output vectors from the $Q$ branches are concatenated along the feature dimension:
\begin{equation}
    {\mathbf{h}_{\rm{merge}}} = \left[ {{\mathbf{h}_1};{\mathbf{h}_2}; \cdots ;{\mathbf{h}_Q}} \right] \in {\mathbb{R}^{\sum\limits_{q = 1}^Q {{M_q}} }}
\end{equation}
where $\left[ { \cdot ; \cdot ; \cdot } \right]$ denotes vector concatenation. The merged vector $\mathbf{h}_{\rm{merge}}$ is then fed into the joint processing layer to obtain:
\begin{equation}\label{gthy}
    \mathbf{z} = \sigma \left( {{\mathbf{W}^\mathbf{z}}{\mathbf{h}_{\rm{merge}}} + {\mathbf{b}^\mathbf{z}}} \right)
\end{equation}
where $\mathbf{W}^\mathbf{z}\in\mathbb{R}^{{\sum\limits_{q = 1}^Q {{M_q}} \times} \frac{1}{2}{\sum\limits_{q = 1}^Q {{M_q}}}}$ is the weight.

Since the MB\_FCNN is designed as a regression network, the final output layer employs a linear activation function. Accordingly, the desired output vector ${{\mathbf{\hat \Theta } }_{d,t}}$ containing the $Q$ true angles is expressed as:
\begin{equation}
  {\widetilde {{\bf{\hat \Theta }}}_{d,t}} = {\mathbf{W}^l}\mathbf{z} + {\mathbf{b}^l},
\end{equation}
where ${\mathbf{W}^l} \in {\mathbb{R}^{\frac{1}{2}\sum\limits_{q = 1}^Q {{M_q}}  \times Q}}$ and $\mathbf{b}^{l} \in \mathbb{R}^{Q}$ denote the weight and bias vector of the output layer, respectively.

Since the observed wavefront characteristics differ across subarray groups, the \( Q \) true direction estimation corresponding to the \( Q \) subarrays are only approximately equal, satisfying:

\begin{align}
{\hat{\theta}_{d,t,1}} \approx {\hat{\theta}_{d,t,2}} \approx  \cdots  \approx {\hat{\theta}_{d,t,Q}} 
\end{align}

Therefore, the predicted vector by MB\_FCNN, \({\widetilde {{\bf{\hat \Theta }}}_{d,t}} = \left[ \widetilde{\hat{\theta}}_{d,t,1}, \widetilde{\hat{\theta}}_{d,t,2}, \cdots, \widetilde{\hat{\theta}}_{d,t,Q} \right]^T\), cannot be directly regarded as the intended DOA measurement. Based on this, an appropriate fusion strategy is required to integrate the \(Q\) estimated DOAs produced by MB\_FCNN and derive a final direction estimation.
Given that a linear mapping theoretically exists between any two numerical values, the fusion strategy can be formulated as a linear combination of the \( Q \) true angle estimation.

\begin{equation}
    \hat \theta  = \sum\limits_{q = 1}^Q {\gamma_q} f_q\left( {{{\widetilde{\hat{\theta}}}_{t,q}}} \right),
\end{equation}  
where \(f_q(\cdot)\) denotes a certain linear mapping function and \(\gamma_q\) represents the corresponding linear combination coefficient. Since a single linear layer (without activation function) in a neural network can accurately represent such linear mappings, a perceptron-based FusionNet is adopted, as illustrated in Fig. \ref{fig_alg_flow5}, with the mapping function formulated as:

\begin{equation}
    \hat \theta  = {f^o}\left( {{{{\widetilde {{\bf{\hat \Theta }}}_{d,t}}}}} \right) = {\mathbf{W}^o}{{{\widetilde {{\bf{\hat \Theta }}}_{d,t}}}} + {\mathbf{b}^o}
\end{equation}
where $\mathbf{W}^o$ and ${\mathbf{b}^o}$ is the weight and bias of fusionNet, respectively. \(\hat{\theta} \) denotes the final estimated DOA value.

Based on the above analysis, the overall mapping relationship of the MBDNN, which is composed of the MB\_FCNN and a single linear layer, can be expressed as:
\begin{equation}\label{hgbbgbn}
    \begin{aligned} 
\hat \theta  &= {f^o}\left( {{f_{\rm{MF}}}({{\bf{x}}_1},{{\bf{x}}_2}, \ldots ,{{\bf{x}}_Q};\mathbb{W})} \right)\\
 &= {f^o}\left( {{f_{\rm{MF}}}\left( {{{{\bf{\hat \Theta }}}_d}\left[ {0:{M_1}} \right], \ldots ,{{{\bf{\hat \Theta }}}_d}\left[ {\sum\limits_{i = 1}^{Q - 1} {{M_i}} :\sum\limits_{i = 1}^Q {{M_i}} } \right]} ;\mathbb{W}\right)} \right)\\
 &= {{\bf{W}}^o}\left( {{{\bf{W}}^l}\left( {\sigma \left( {{{\bf{W}}^{\bf{z}}}\left[ {{{\bf{h}}_1};{{\bf{h}}_2}; \cdots ;{{\bf{h}}_Q}} \right] + {{\bf{b}}^{\bf{z}}}} \right)} \right) + {{\bf{b}}^l}} \right) + {{\bf{b}}^o}
\end{aligned}
\end{equation}

\subsection{Training loss function}
In the regression task, the MSE is adopted as the loss function, and the Adam optimizer is used for training. For the training dataset \(\mathbb{T}_{{\rm{MB\_FCNN}}}\), the loss function is constructed as:  

\begin{equation}
 \begin{aligned}
{L_{\rm{MB\_FCNN}}} &= \frac{1}{{{Q{\rm{N}}_{{\rm{total}}}}}}\sum\limits_{d = 1}^{{{\rm{N}}_{{\rm{total}}}}} {\left\| {{{{\bf{\hat \Theta }}}_{d,t}} - {{\widetilde {{\bf{\hat \Theta }}}}_{d,t}}} \right\|_F^2} \\
 &= \frac{1}{{{Q{\rm{N}}_{{\rm{total}}}}}}\sum\limits_{d = 1}^{{{\rm{N}}_{{\rm{total}}}}} {\sum\limits_{q = 1}^Q {{{\left( {{{\hat \theta }_{d,t,q}} - {{\widetilde {\hat \theta }}_{d,t,q}}} \right)}^2}} } 
\end{aligned}
\end{equation}
where ${{\rm{N}}_{{\rm{total}}}}$ denotes the total number of training samples, \(\widetilde {{\bf{\hat \Theta }}}_{d,t}\) is the predicted DOA vector of the MB\_FCNN for the \(d\)-th sample, and \(\mathbf{\hat{\Theta}}_{d,t}\)is the corresponding truth label.

Similarly, the loss function for \(\mathbb{T}_{{\rm{FusionNet}}}\) is defined as:
\begin{equation}
   {L_{{\rm{FusionNet}}}} = \frac{1}{{{{\rm{N}}_{{\rm{total}}}}}}\sum\limits_{d = 1}^{{{\rm{N}}_{{\rm{total}}}}} {\left\| {\hat \theta  - \widetilde {\hat \theta }} \right\|_F^2} 
\end{equation}

The overall training loss of the MBDNN is defined as:
\begin{equation}
   {L_{{\rm{MBNN}}}} = \frac{1}{{{Q}{{\rm{N}}_{{\rm{total}}}}}}\sum\limits_{d = 1}^{{{\rm{N}}_{{\rm{total}}}}} {\sum\limits_{q = 1}^Q {{{\left( {{{\widetilde {\hat \theta }}_{d,t,q}} - \hat \theta } \right)}^2}} } 
\end{equation}
Thus, the training objective is to minimize the ${L_{{\rm{MBNN}}}}$.

\section{Simulation Results}\label{sec_simu}
This section provides a set of experimental results to evaluate the effectiveness of the two introduced DOA sensing approaches: CRLB-ratio-WF and MBDNN.
In addition, the root-mean-squared error (RMSE) is adopted as the evaluation metric for DOA estimation accuracy, with the CRLB of the H$^2$AD array serving as the performance benchmark. The RMSE is formulated as:
\begin{align} \label{}
	RMSE = \sqrt {\frac{1}{L}\sum_{l}^{L}(\hat{\theta}_{l}-\theta_0)^2}
\end{align}
where $\hat{\theta}_{l}$ denotes the estimated DOA obtained from the $l$-th Monte Carlo simulation trial, $\theta_0$ represents the true DOA of the signal source (i.e., the actual incident angle). $L$ is the number of Monte Carlo simulation trials. And Table \ref{tableDOA} summarizes the key system parameters involved in the simulation.

\begin{table}[http]
\caption{System model parameters}
\label{tableDOA}
\centering
\begin{tabular}{ccl}
\cline{1-2}
Parameters  & Values &  \\ 
\cline{1-2}
Groups in $\rm{H}^2$AD: $Q$ & 3 &\\
Antennas in each group: $M_1, M_2, M_3$ & 7, 11, 13 &  \\
Subarrays in each group: $K_1, K_2, K_3$ & 16, 16, 16 &\\
Transmitter angle: $\theta_0$ & 41$^\circ$  &  \\
Number of snapshots: $T$& 200 &\\
Monte Carlo experiments: $L$ & 5000 &  \\
DOA step size: $s$     &1$^\circ$ &  \\
DOA search range: $[\theta_{\rm{min}},\theta_{\rm{max}}]$ & [-90$^\circ$, 90$^\circ$] &\\
SNR range: $[\rm{SNR}_{min},\rm{SNR}_{max}]$  & \{-15:5:15\}dB &\\
Learning rate &  $1 \times {10^{ - 4}}$ &\\
Epoch & 100 &\\
\cline{1-2}
\end{tabular}
\end{table}

\begin{figure}[!http]
	\centerline{\includegraphics[width=3.5in]{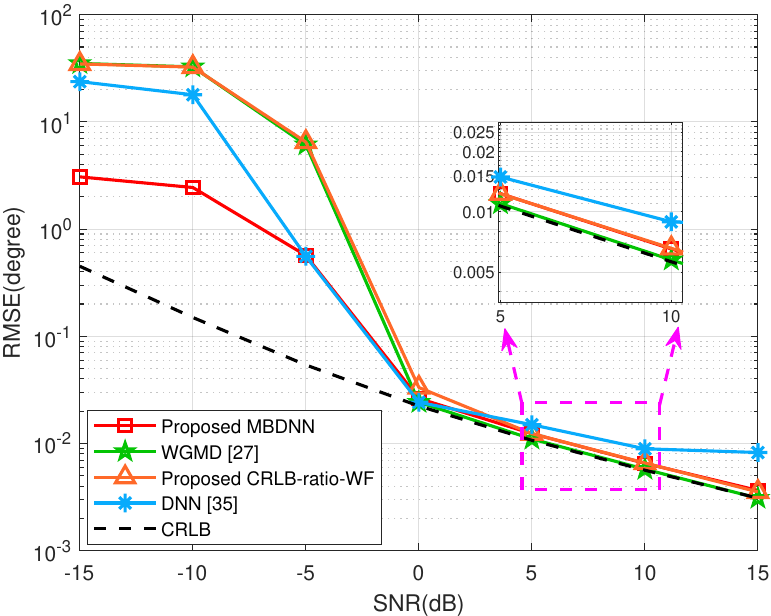}}
	\caption{RMSE versus SNR for the proposed and compared methods.\label{RMSE_SNR}}
\end{figure}

Fig. \ref{RMSE_SNR} presents the RMSE versus SNR for the proposed and baseline methods (e.g., WGMD \cite{10767772}, DNN \cite{10538312}) with snapshots $T = 100$. As the SNR increases, the RMSE of all methods approaches the CRLB around SNR $ = 0$ dB. Notably, the proposed MBDNN method exhibits significant performance advantages under SNR $\le 0$. In particular, at SNR $= -15$ dB, its DOA sensing accuracy is approximately ten times that of the other methods. Furthermore, although the proposed CRLB-ratio-WF method performs slightly worse than the WGMD method, it relies on far less prior information, making it more practical and adaptable in real-world applications.

\begin{figure}[!http]
	\centerline{\includegraphics[width=3.55in]{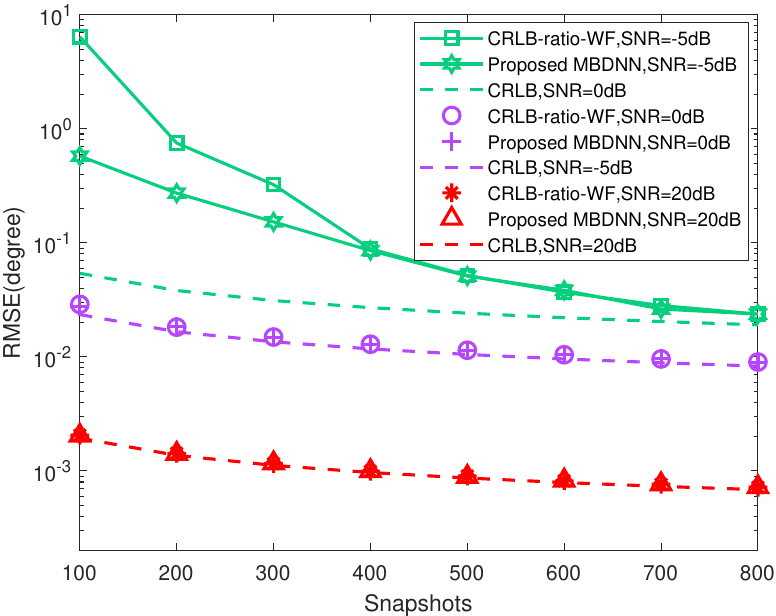}}
	\caption{RMSE versus number of snapshots for the proposed methods.\label{RMSE_snap}}
\end{figure}

Fig. \ref{RMSE_snap} illustrates the relationship between the RMSE and the number of snapshots under three different SNR (SNR$\in\{-5,0,15\}$dB). 
At low SNR ($-5$ dB), the two methods show significant improvement as the number of snapshots increases, with the MBDNN method exhibiting greater robustness.
At medium-high SNR (e.g., SNR $\ge 0$ dB), the RMSEs of the proposed methods converge more rapidly and closely approach the CRLB. As the SNR increases, the RMSE becomes increasingly insensitive to the number of snapshots, achieving near-CRLB performance even with a small number of snapshots.

\begin{figure}[!http]
	\centerline{\includegraphics[width=3.5in]{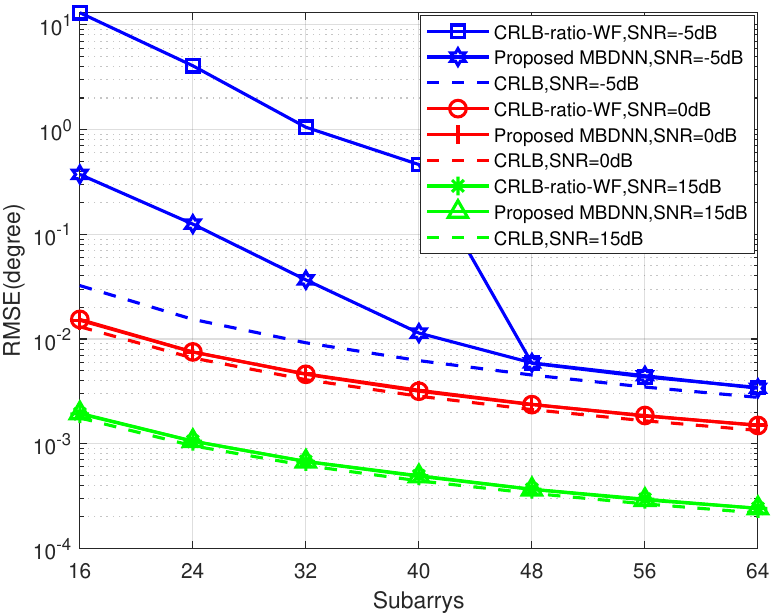}}
	\caption{RMSE versus number of subarrays for the proposed methods.\label{RMSE_an}}
\end{figure}

Fig. \ref{RMSE_an} presents RMSE versus the number of subarrays with $T = 200$. In the experiment, the subarray antenna 
configurations were fixed as $M1=11$, $M2=13$, $M3=17$, and the number of subarrays in each group was set equal ($K1 = K2 =K3$), where $K_q$ increases from $16$ to $64$  an interval of $8$. The estimation accuracy improves with increasing $K_q$. At SNR $\ge 0$ dB, the DOA estimation accuracy of two proposed methods closely approaches the CRLB. Taken together with Figs \ref{RMSE_SNR} and \ref{RMSE_snap}, the proposed methods exhibit excellent DOA estimation performance in medium-to-high SNR regime.

\begin{figure}[!http]
	\centerline{\includegraphics[width=3.5in]{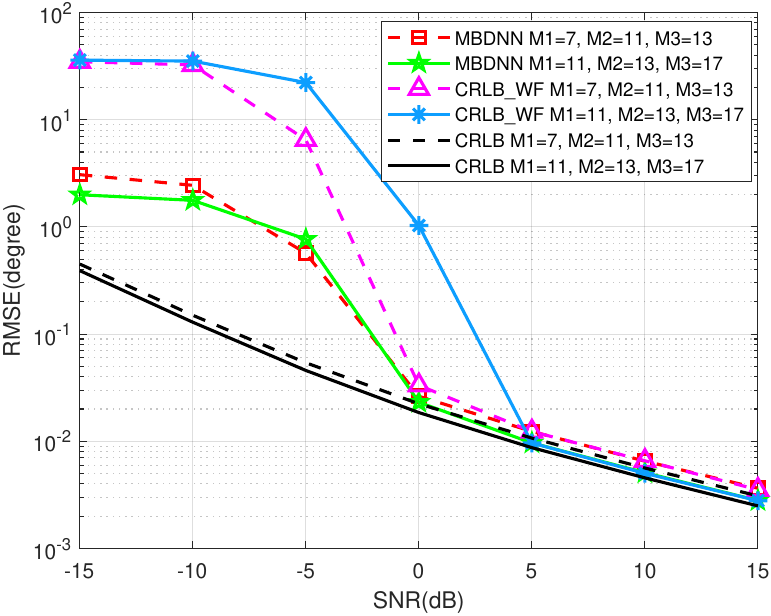}}
	\caption{RMSE versus SNR for different number of sub-array antennas.\label{RMSE_SNR_M}}
\end{figure}

Fig. \ref{RMSE_SNR_M} illustrates the RMSE versus SNR of the proposed methods under different subarray antenna configurations 
($M1, M2, M3$). It can be observed that the RMSE of all methods decreases with increasing SNR and approaches the CRLB in medium-high SNR regions. Notably, as the number of subarray antennas increases, the CRLB decreases, further validating the proportional relationship described by Eq. (\ref{defrfgrgr}). Furthermore, larger subarray configurations (e.g., $M1=11, M2=13, M3=17$) result in lower RMSE values in the medium-high SNR range. Both the MBDNN method and the CRLB-ratio-WF method approach the CRLB.
 
\begin{figure}[!http]
	\centerline{\includegraphics[width=3.5in]{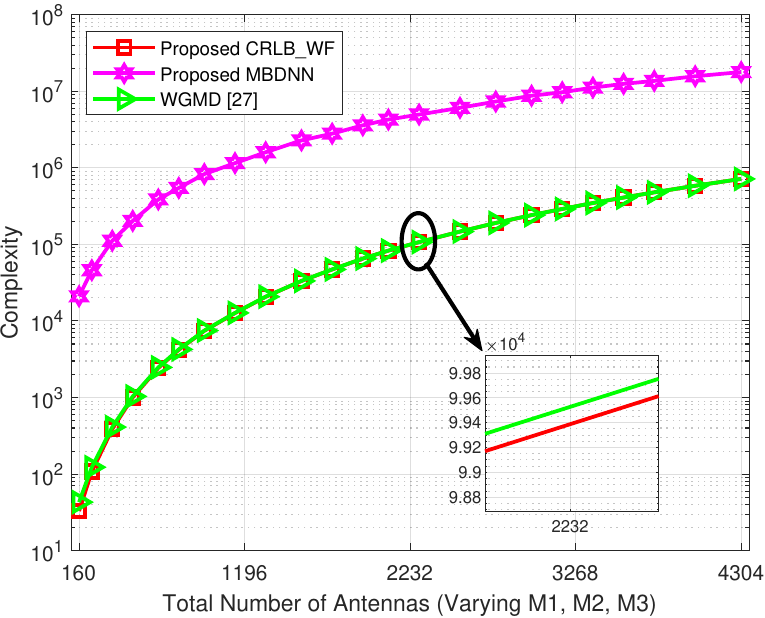}}
	\caption{Computational complexity for the proposed methods with varying M1, M2, M3.\label{complexity}}
\end{figure}

The two proposed methods exhibit identical computational complexity during the DOA sensing stage. Therefore, disregarding its computational burden, the trends in computational complexity versus the total number of antennas under varying$M_1$, $M_2$, and $M_3$, are illustrated in Fig. \ref{complexity}. As the input vector size of the MBDNN is $M_s$, and its network depth is strongly correlated with the input size, an increase in $M_s$ leads to a corresponding rise in the MBDNN’s complexity. In contrast, the CRLB-ratio-WF method does not require prior variance estimation and relies on less input information, resulting in slightly lower complexity compared to the WGMD method.

\section{Conclusion}
This paper proposed two innovative approaches, a low-complexity CRLB-ratio-WF method and a MBDNN architecture for enhanced DOA sensing, 
to address the challenges of high computational complexity and heavy prior reliance in DOA estimation for massive H$^2$AD MIMO receive arrays.
The proposed CRLB-ratio-WF method approximated the inverse of the CRLB for each subarray using the reciprocal of its antenna number, eliminating the need for explicit CRLB computation and dramatically requiring less prior information. 
The MBDNN effectively eliminated false solutions from the DOA candidate sets of all the subarray and fused the inferred true angles via a shared regression module, obtaining the enhanced final DOA sensing. 
Experimental results demonstrated that the proposed CRLB-ratio-WF method achieved DOA estimation performance comparable to traditional CRLB-based approach with a significant reduction in the reliance on prior knowledge, while the MBDNN provided superior estimation accuracy in low-SNR regions, achieving a tenfold improvement over the CRLB-ratio-WF method at SNR = $-15$ dB.
The two proposed methods provide a comprehensive solution for DOA sensing across different scenarios: the CRLB-ratio-WF method is well-suited for conventional environments with limited computational resources, while the MBDNN offers a promising solution for high-accuracy requirements under low-SNR conditions.

\vfill\pagebreak

\end{document}